\documentclass{easychair}

\usepackage{doc}

\usepackage{doc}
\usepackage{multirow}

\title{Towards an Automated Pipeline for Detecting and Classifying Malware through Machine Learning}

\author{
	Nicola Loi\inst{1}
	\and
	Claudio Borile\inst{2}
	\and
	Daniele Ucci\inst{2}
}

\institute{
	Università degli Studi di Torino,
	via Pietro Giuria 1, 10125 Turin, Italy\\
	\email{[name].[surname]@edu.unito.it}
	\and
	aizoOn Technology Consulting, Strada del Lionetto 6, 10146 Turin, Italy\\
	\email{[name].[surname]@aizoongroup.com}
}

\authorrunning{Loi, Borile and Ucci}

\titlerunning{Malware classification pipeline}

\begin{document}

\maketitle

\begin{abstract}
The constant growth in the number of malware - software or code fragment potentially harmful for computers and information networks - and the use of sophisticated evasion and obfuscation techniques have seriously hindered classic signature-based approaches.
On the other hand, malware detection systems based on machine learning techniques started offering a promising alternative to standard approaches, drastically reducing analysis time and turning out to be more robust against evasion and obfuscation techniques.  
In this paper, we propose a malware taxonomic classification pipeline able to classify Windows Portable Executable files (PEs). Given an input PE sample, it is first classified as either malicious or benign. If malicious, the pipeline further analyzes it in order to establish its threat type, family, and behavior(s).
We tested the proposed pipeline on the open source dataset EMBER, containing approximately $1$ million PE samples, analyzed through static analysis.
Obtained malware detection results are comparable to other academic works in the current state of art and, in addition, we provide an in-depth classification of malicious samples. Models used in the pipeline provides interpretable results which can help security analysts in better understanding decisions taken by the automated pipeline.

\end{abstract}

\section{Introduction}
\label{sec:intro}

Malware (short for malicious software) is the generic term used to refer to unwanted software developed to infect and interfere with the operations of a single machine or networks of computers \cite{sikorski2012practical,milovsevic2013}. Since the first documented virus appeared in the 1970s, the evolution of computer science has always been accompanied closely by the creation of new, better and more harmful malicious software, in a constant fight between malware developers and security analysts. In recent years, though, the refinement and emergence of new software technologies have allowed an exponential growth in the number of malware in circulation, not only vertically (volume) but also horizontally (types and functionality) \cite{avtest2021}. Together with the ever more sophisticated evasion techniques being developed by attackers, security experts and anti-malware vendors struggle to keep up the race by means of ``standard" methods, i.e. signature-based and heuristics \cite{sihwail2018survey,jawad2019n,monnappa2018learning}. In this context, machine learning (ML) seems to be the most promising tool for an automated analysis and prevention of this kind of threats \cite{souri2018state}.
The strength of ML is its ability to automatically identify hidden patterns and correlations in large volumes of raw data, and exploit these statistical features to, in the case of malware analysis, recognise previously unseen attacks. Generally speaking, classic ML approaches for cyber security purposes focus on a first phase of features extraction through static, dynamic or hybrid analysis. These features are then used to train models that allow to classify malicious and benign files. More recently, the advancements in deep learning methods has inspired a series of studies exploiting other input formats, like raw binaries \cite{rafique2019,raff2017} and image representations \cite{ahmadi2016,nataraj2011malware,ghouti2019malware} among others. Historically, researchers and security vendors have mostly been more focused on creating models for the detection of malicious and benign files rather than exploring the possibility of using ML for an in-depth analysis of single malware samples. A reason for this might be the difficulty in collecting large and well-annotated datasets, a complex and expensive task, together with the intrinsic difficulty in classifying the characteristics of  malwares due to the presence of many variants and the lack of a standard nomenclature \cite{maggi2011finding,szor2005art,ducau2019automatic}. 

Endgame --- now Elastic --- released in 2017 the first version of an open-source dataset called EMBER (Elastic Malware Benchmark for Empowering Researchers) containing semi-raw static features from more than $1$ million Portable Executable (PE). In 2018, they released a second version of the dataset, using more recent malware, correcting issues in the data collection, and providing also labels for malware classes, using the open source tool AVClass~\cite{avclass1}. This tool allows to parse and organise, in a definite taxonomy, the different nomenclatures returned by multiple anti-virus vendors. Exploiting and integrating the last version of the EMBER dataset as described in detail below, we explore the possibility of training a machine learning pipeline for a full, automated analysis of PEs using static features, from malware detection to classification by threat-type, family, and behaviour.
At the time of writing, there is no academic paper that has explored the possibility of leveraging the EMBER dataset for a taxonomic malware classification. The paper is organized as follows: Section~\ref{sec:related} presents related works, while Section~\ref{sec:ember} details the EMBER dataset. The proposed classification pipeline is described in Section~\ref{sec:pip} and experimental evaluation results are reported in~\ref{sec:pipe_acc}.
\section{Related Work}
\label{sec:related}

In the last decade, the number of studies on machine learning detection techniques is constantly increased thanks to: i) the recent growth of new and powerful algorithms and data wrangling methods, ii) the increase in computational capabilities, and iii) the availability of public, annotated malware datasets~\cite{gibert2020,ucci2019,rahul2020, singh2020,raff2020}. However, most of these works only consider the problem of distinguishing malicious software from benign. Furthermore, many papers rely on small or outdated datasets that are unlikely to provide a statistically-significant representation of malware population and labelling procedures create a bias towards easier datasets \cite{smith2020,fuyong2017}.
The above considerations resulted in a general difficulty in assessing the real performance of these methods ``into the wild" and a general incapacity to deliver models that can be effectively deployed, despite notable results summarized in~\cite{smith2020}.
Nevertheless, many recent works have been very effective in detecting malware, reaching almost perfect performances in accuracy.

In this scenario, the authors of the EMBER dataset provided a benchmark model in 2018, trained on their latest release of the model, obtaining a 86.8\% detection rate at 0.1\% False Positive Rate (FPR).
Results obtained in the previous model release reported a 93\% detection rate~\cite{roth2019}, meaning that the EMBER dataset developers have successfully hardened the process of correctly classifying malicious samples.
Despite more recent works on the same dataset provide small improvements in the detection rate~\cite{oyama2019,pham2018}, they usually rely on deep learning frameworks that make more difficult to interpret model outputs. 

To our knowledge, there are no published works that used the 2018 EMBER dataset for a multi-class classification in malware families. Among the studies that tackle the problem of classifying malware families \cite{rieck2008,nassar2019,hu2013mutantx,sun2017malware}, Ahmadi \emph{et al.} \cite{ahmadi2016} considered the Microsoft Malware Classification Challenge data, a labeled dataset of about $20,000$ malware samples representing a mix of 9 different families\footnote{https://www.kaggle.com/c/malware-classification}, to build a model for malware families classification. While the dataset is much smaller than EMBER and thus the performances are difficult to compare, they obtained a notable $0.997$ accuracy over a $5$-fold cross validation, considering a set of features similar to those used in the EMBER model, suggesting their robustness for both malware detection and family classification.

Malware behaviour classification is usually based on dynamic or hybrid analysis~\cite{yu2018, brumley2008automatically,bekerman2015unknown}. This is not surprising, since malware behaviour is expected to manifest itself only upon execution~\cite{smith2020}. Nevertheless, we believe that it is still interesting to assess behavioral classification on purely static analyses contained in EMBER. We have not find any previous work considering the problem of an integrated, hierarchical  malware classification into threat-type, family, and behaviour using features from static analysis.
\section{Data description}
\label{sec:ember}

\subsection{The EMBER dataset}
In this work we consider the 2018 release of the EMBER (Elastic Malware Benchmark for Empowering Researchers) dataset, an open source benchmark collection of 1 million PEs scanned in or before 2018 \cite{ember2018}. The data comes split in two separate sets containing the training and test data. The training set consists of 600,000 labeled samples (benign or malware) and 200,000 unlabeled samples the we did not consider in this study. The test set contains 200,000 labeled samples. The authors claim that the particular splitting between train and test is specifically engineered for having a ``harder" dataset with respect to the first release \cite{roth2019}. Each sample comes as a JSON object and is uniquely identified via its sha256 and md5 hashes, and provides semi-raw information for static malware analysis parsed using the LIEF open source package \cite{lief} and divided in nine major groups: General, header, and section information, imported and exported functions, strings information, raw-byte histogram, and byte-entropy histogram. Finally, additional information is given on the label (0 for benign files, 1 for malicious files, and -1 for unlabeled files), the coarse time stamp of estimate first detection of the malware, the malware class extracted from the VirusTotal \cite{virustotal} report using the open source tool avclass \cite{avclass1}.
A full description of the dataset can be found in \cite{ember2018, roth2019}. As explained in detail in section ~\ref{sec:pip}, we consider all the 600,000 labeled training data for training and validating each stage of the classification pipeline, while the test set will be used only to asses the global performance of the pipeline, mimicking its usage in a real-world scenario.

\subsection{Threat-type and behaviour labels collection}
\label{sec:labelling}
The EMBER dataset only provides the malware/benign and a generic ``avclass'' labels (the output of the AVClass tool \cite{avclass1} for family tagging). In order to classify a specific malware sample by threat-type, family, and behavior, we used the open-source tool AVClass2 \cite{avclass2}, not yet available when the EMBER dataset came out, to systematically extract information on the malware threat-type and behavior for each labeled sample in the dataset. Our final dataset consists of the data from the EMBER original dataset plus four labels: malware/not-malware, and if malware its threat-type, family and behavior classes where available. It is important to note that AVClass2 parses the malware label returned from each of the AV vendors and returns a ranking of the returned labels, not necessarily a single result per class. As a consequence, while a powerful tool tailored for each AV vendor and including a complete malware taxonomy, still suffers from the lack of a shared nomenclature across different security vendors. Furthermore, the estimated accuracy of the output labels is reported to be around 90\% for families (threat-type and behavior not reported) \cite{avclass2}, so that we must take into consideration the inherent imperfect labelling of the data.

\subsection{Classification target}
\label{sec:clf_targ}

\begin{figure}[tp]
	\centering
	\begin{minipage}{0.33\textwidth}
		\centering
		\includegraphics[height=6cm, width=\textwidth]{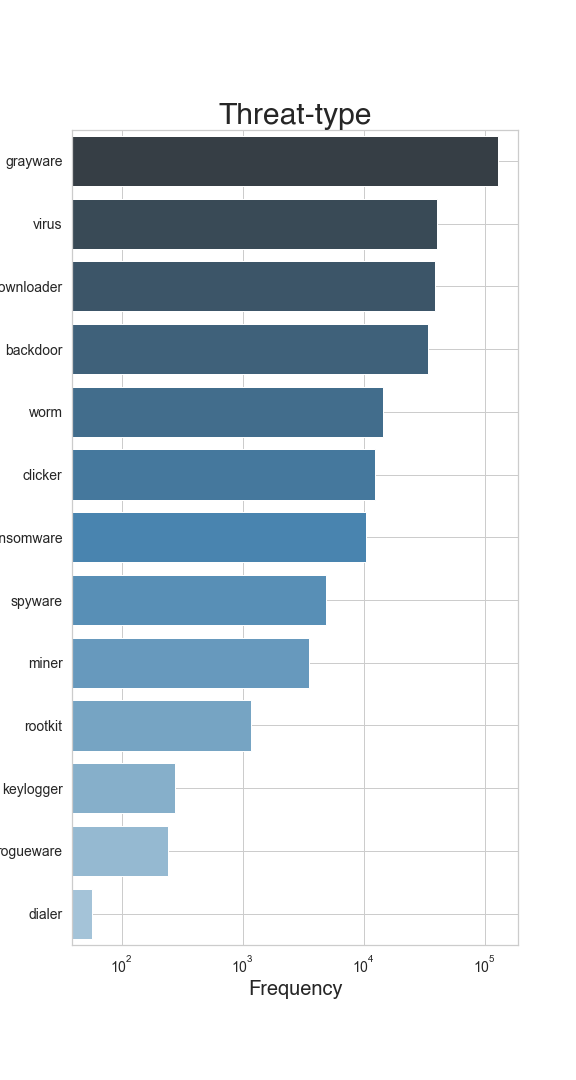}
	\end{minipage}
	\begin{minipage}{0.33\textwidth}
		\centering
		\includegraphics[height=6cm, width=\textwidth]{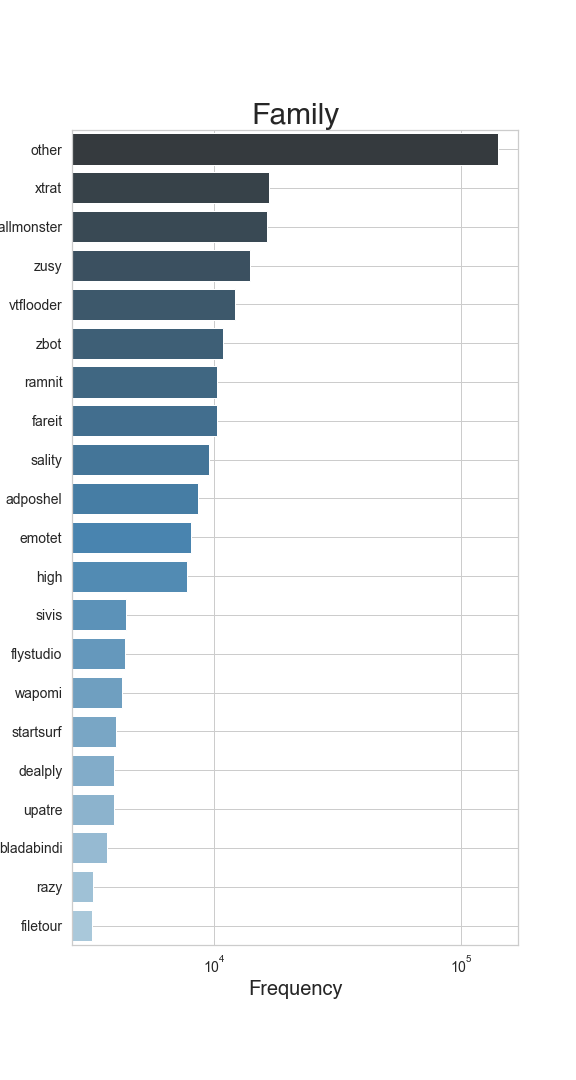}
	\end{minipage}\hfill
	\begin{minipage}{0.33\textwidth}
		\centering
		\includegraphics[height=6cm, width=\textwidth]{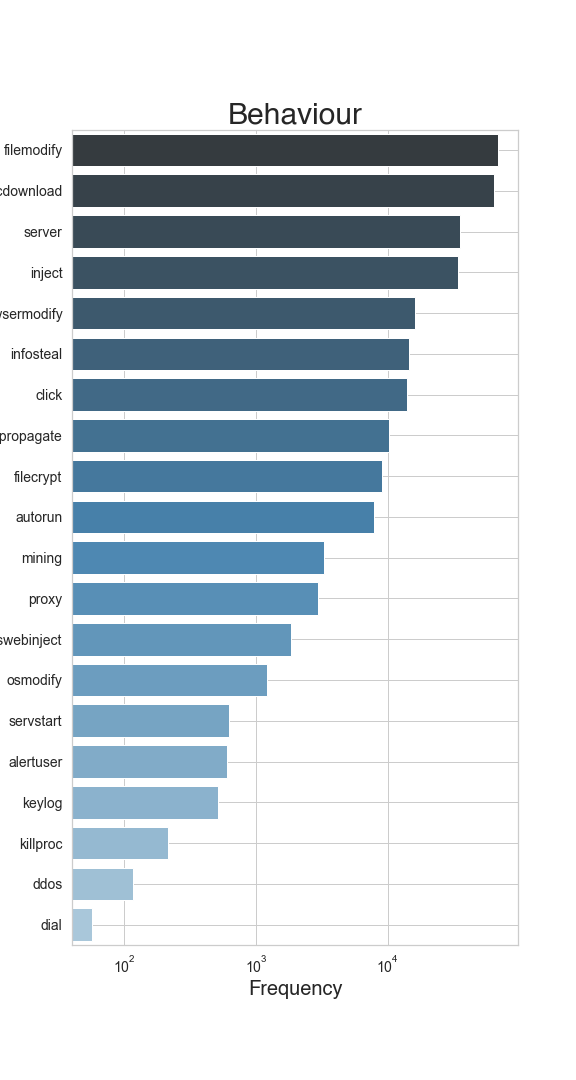}
	\end{minipage}
	\caption{Frequency of the target classes for the three different types of classification stages. The horizontal axis is in logarithmic scale.}
	\label{fig:numerosity_train}
\end{figure}

Looking at figure \ref{fig:numerosity_train} we can see that for each of the classification stages of the pipeline the classes are heavily unbalanced (note the logarithmic scale). In particular, the ``Family" group has a very large number of poorly populated classes in the train set. We thus kept the first 20 families and grouped all the remaining in a single class named ``other", that now represents the most populous class. We will discuss the effects of this unbalance in section \ref{sec:pipe_acc}. For the threat-type we can see that, as expected, the most represented classes are greyware, viruses, and downloaders, while dialers are by far the less common with only few tens of samples. Regarding malware behavior, the first two classes -filemodify and execdownload- make half of the total samples. Coherently with the threats, class ``dial" is the class with fewest samples.

\section{Pipeline Architecture Overview}
\label{sec:pip}

\begin{figure}[tp]
	\label{fig:pipeline}
	\centering
	\includegraphics[width=0.75\textwidth]{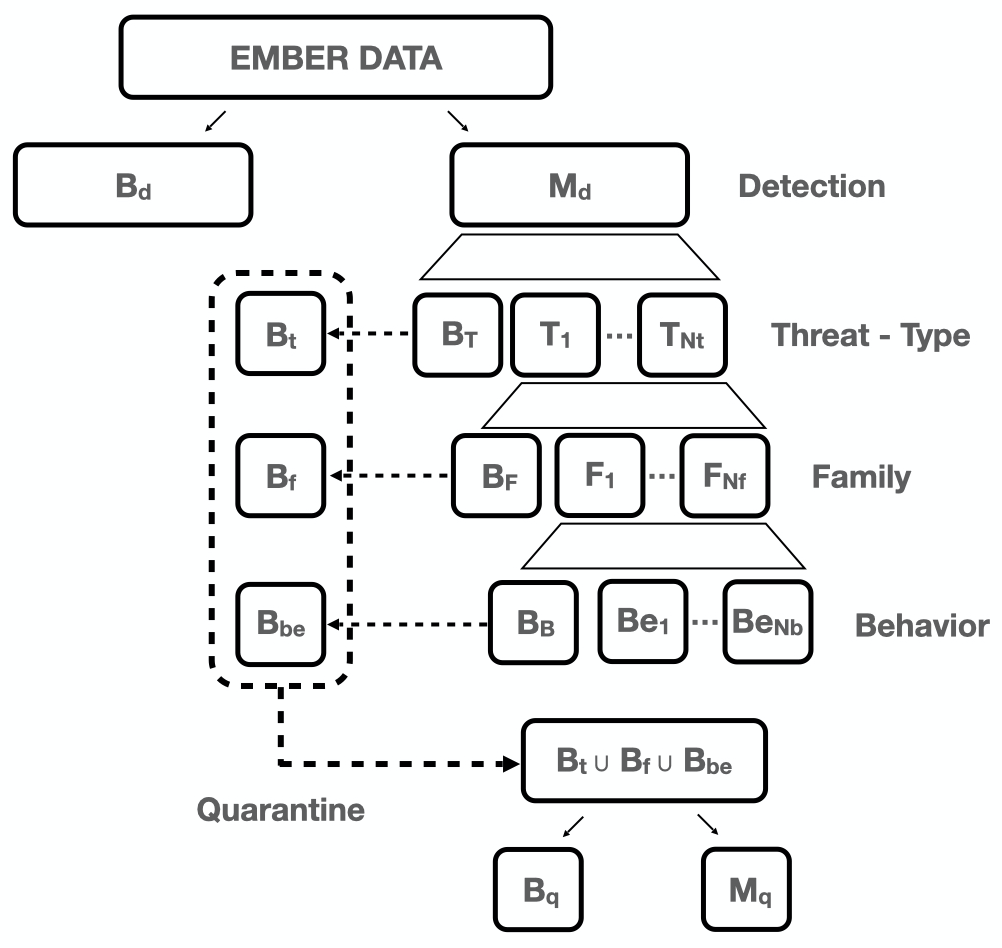}
	\caption{Representation of the proposed classification pipeline.}
\end{figure}

As introduced in Section~\ref{sec:intro}, the main goal of the proposed pipeline is to accurately classify malware using different stages of processing, each one leveraging a properly trained classifier.
Figure~\ref{fig:pipeline} shows how we implemented the pipeline: the first stage detects whether input samples are malicious or benign; those classified as malicious are propagated to the second stage, which is responsible for labeling them with a known malware category (e.g., virus, backdoor, greyware). 
It is worth noting that, in this and the next stages, we employ (different) classifiers able to recognize misclassified benign samples, results of errors occurred in the previous stages. After being categorized, malicious files are further classified in families (e.g., \textit{xtrat} and \textit{vtflooder}). 
Analogously to the previous stage, the employed classifier is able to discriminate benign samples --- identified as malicious --- but also classify malicious samples belonging to less recurrent families into a specific class, called \textit{other}. 
Finally, different malware families are classified according their malicious behaviour.
Whenever a sample is classified incoherently through the pipeline, that is, is classified as malware is the detection stage but as a benign in some next stage, we envision another stage that gathers all these suspicious samples, the \emph{quarantine}, and reclassifies again them to improve the overall malware detection performance.

For training pipeline's classifiers, we use the same data flow described above: samples used to validate a classifier in a stage are, then, provided in input to the next stage to train the following classifier.
During the training phase, at each stage of the pipeline -detection and threat-type, family, and behavior classification-, the output of the preceding stage is split into train and validation sets, and a Gradient Boosting Decison Tree (GBDT) classifier is trained and its hyper-parameters optimized for the specific task. We used the Python API implementation of Microsoft's LightGBM\footnote{https://github.com/microsoft/LightGBM} open source framework, that proved to be the the optimal solution between training time and performance \cite{galen2020}.

As described in the original EMBER benchmark model \cite{roth2019}, the raw features of each sample are mapped into a fixed-size vector of length 2351. In this work we start from the same feature set but we one slight modification. The  original  feature  vectors  rely  heavily  on  the  hashing trick in order to contain the variable and potentially very large number (order of millions) of imported functions,  while here we chose to keep only the first 151 most common ports in order to keep the balance with the other major groups of features. A comparison between our model and the original EMBER results shows that the model performance is comparable (86.3\% TPR @ 0.1\% FPR versus 86.8\% TPR @ 0.1\% FPR of EMBER's model), but with the advantage of a more interpretable result in our case.

\section{Experimental Evaluation}
\label{sec:pipe_acc}

\begin{table}[bp]
	\footnotesize
	\caption{Results of the experimental evaluation carried out on the EMBER dataset reporting accuracy, AUC, false positives, and false negatives metrics both for validation and test phases.}
	\begin{center}
		\begin{tabular}{c c c c c | c c c c}
			\cline{2-9}
			& \multicolumn{4}{  c | }{Validation} & \multicolumn{4}{ c }{Test}\\
			\cline{1-9}
			Metrics & Detection & Type & Family & Behavior & Detection & Type & Family & Behavior\\ \cline{1-9}
			Samples & 300,000 & 72,257 & 35,934 & 17,842 & 200,000 & 99,314 & 98,611 & 98,452\\
			Accuracy & 0.981 & 0.893 & 0.891 & 0.841 & 0.969 & 0.847 & 0.890 & 0.837\\
			AUC & 0.997 & 0.981 & 0.988 & 0.972 & 0.995 & 0.961 & 0.984 & 0.952\\
			False positives & 2,564 & 1,030 & 482 & 214 & 3,136 & 2,945 & 2,853 & 2,740\\
			False negatives & 3,225 & 202 & 78 & 70 & 3,033 & 512 & 67 & 181\\
			\cline{1-9}
		\end{tabular}
	\end{center}
	\label{tbl:pipe_acc}
\end{table}

Table~\ref{tbl:pipe_acc} summarizes the experimental evaluations carried out on the proposed pipeline. For completeness, it reports both the results obtained during the validation and testing of the classifiers.
Obtained results in the first stage (i.e., detection stage) are comparable with those obtained by EMBER developers~\cite{ember2018}. It is important to note that our pipeline is able to correct false positives (benign samples detected as malicious), as discussed later in this section.

\begin{figure}[h!]
	\centering
	\begin{minipage}{0.49\textwidth}
		\flushleft
		\includegraphics[clip, trim=0.5cm 0cm 0.5cm 0cm, width=1.3\textwidth]{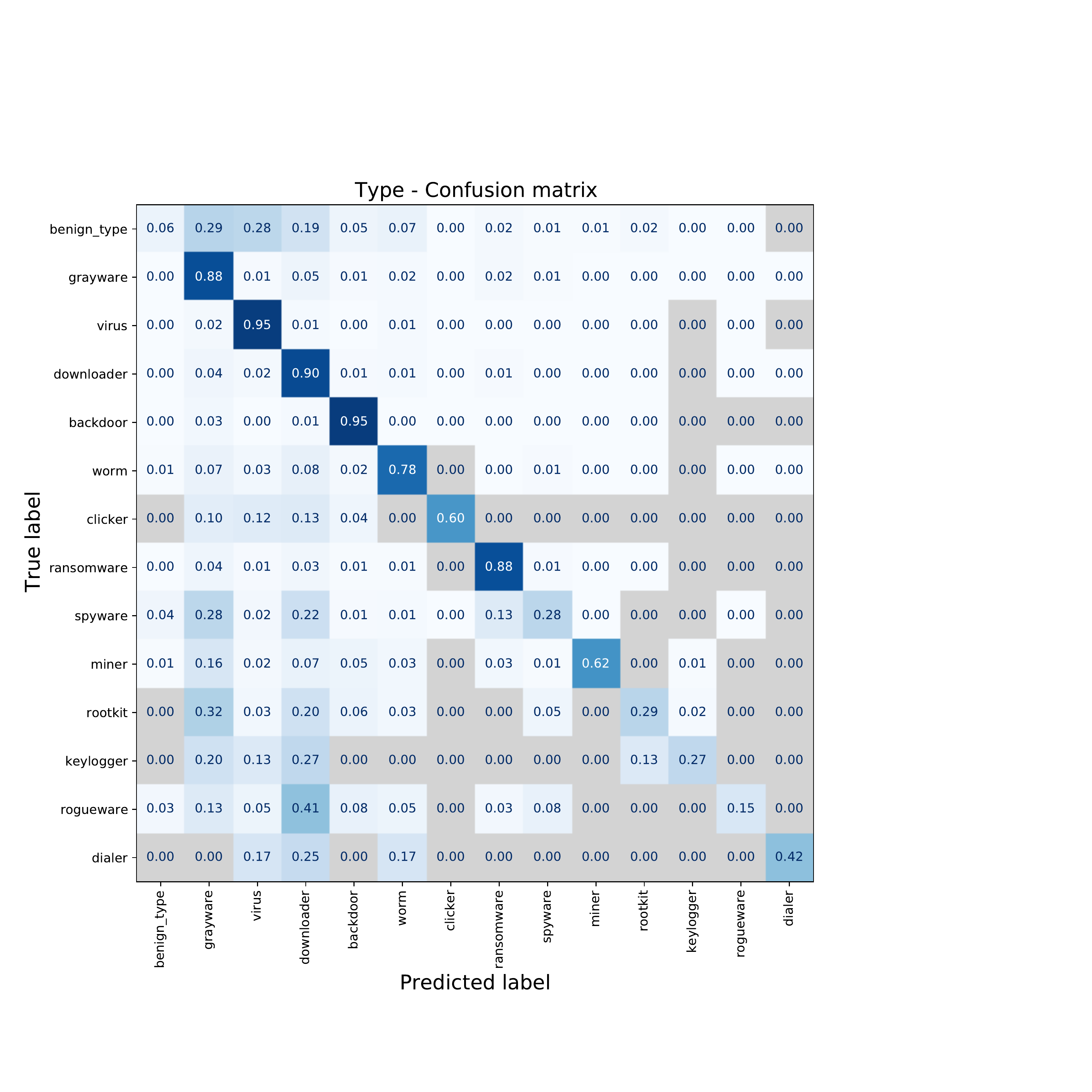}
		\caption{Confusion matrix for the threat-type classification stage reporting the performance, in terms of accuracy, on the test set described in Table~\ref{tbl:pipe_acc}.}
		\label{fig:type_cm}	
	\end{minipage}\hfill
	\begin{minipage}{0.49\textwidth}
		\flushright
		\includegraphics[clip, trim=0.5cm 0cm 0.5cm 0cm, width=1.3\textwidth]{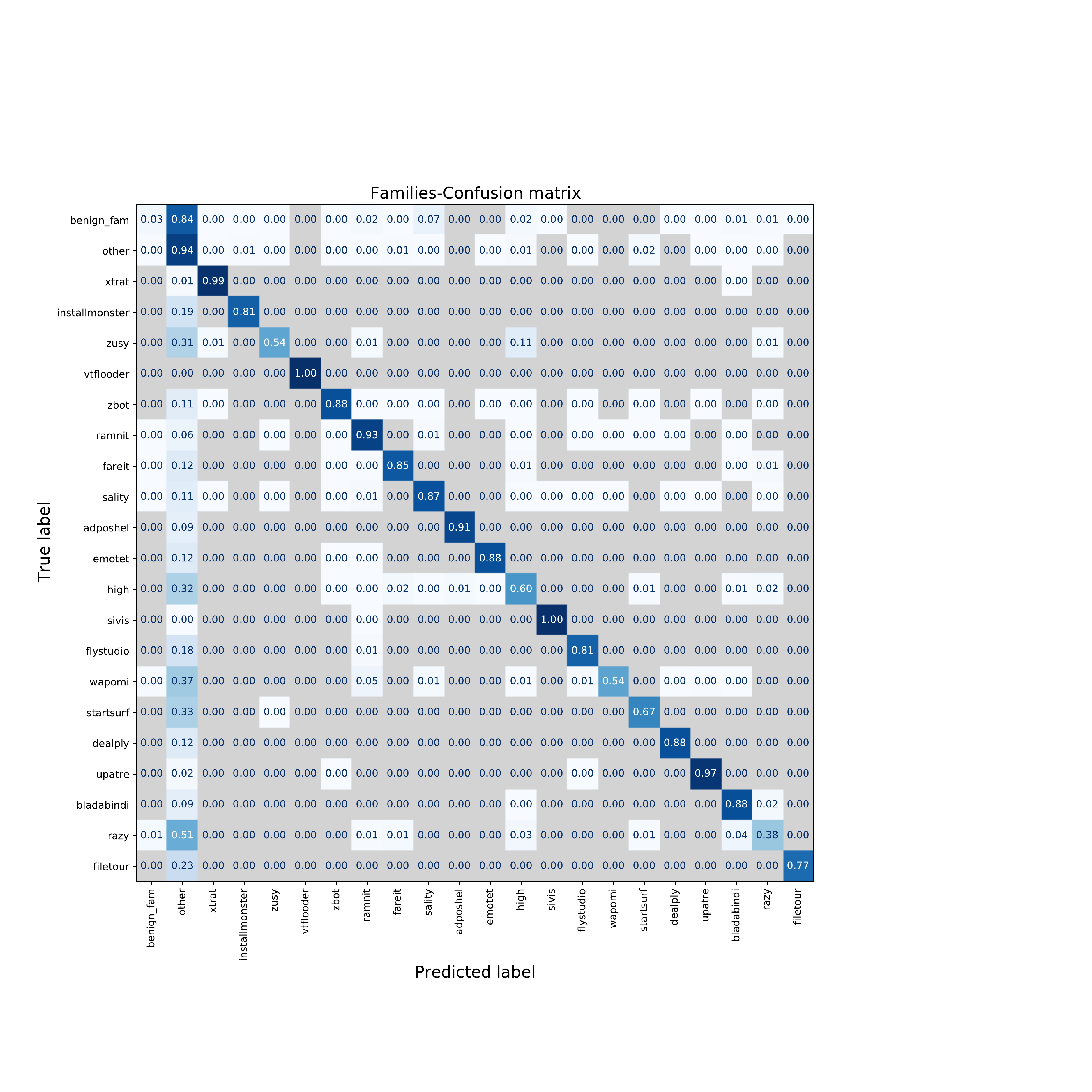}
		\caption{Confusion matrix for the family classification stage reporting the performance, in terms of accuracy, on the test set described in Table~\ref{tbl:pipe_acc}.}
		\label{fig:fam_cm}
	\end{minipage}
\end{figure}

Samples predicted as malicious are then propagated to the stage that classifies threat types.
Results reported in Figure~\ref{fig:type_cm} show that this stage has encountered the most troubles in classifying samples: indeed, benign samples are easily confused with grayware, viruses, and downloaders and the separation among different classes is not so clear (for example, as rootkits and grayware). 
Another aspect to take into account is the fallacy of the ground truth we used to train the threat-type stage classifier: further discussions are outlined in Section~\ref{sec:discussion}.
Conversely to the previous stage, family classifier is more accurate in classifying specific families.
More than $63\%$ of families have been correctly classified with an accuracy greater than $85\%$, with many families having almost all their test samples identified by our classifier (e.g., \textit{xtrat}, \textit{vtflooder}, \textit{sivis}, and \textit{upatre}).
$84\%$ of misclassified benign samples fall in the ‘other’ class and this can be explained by having too few benign samples to build a classifier that is able to identify them.
Further discussions are left in Section~\ref{sec:discussion}.

The last stage of the pipeline consists in classifying malicious samples by their behavior. 
For space constraints, we do not show the confusion matrix associated to the behavior stage.
Similar to the threat-type classification stage, the behavior classifier fails to correctly recognize some specific behaviors (such as, \textit{autorun} and \textit{osmodify}). In addition to potential similar behaviors, another cause of misclassification is the few number of samples employed to train the behavior classifier: less than $20,000$ instances to train the classifier on more than $20$ different behaviors.

As described in Section~\ref{sec:pip}, samples classified as benign in stages after the first one are quarantined for further analyses.
Of all the 200,000 total samples in the test set, only 1150 (0.57\%) end up in the quarantine stage. The classifier is able to recover 221 benign samples that were misclassified in the initial detection stage and, more importantly, recovered 635 malicious samples that were incorrectly labeled as benign in one of the classification stages.

\subsection{Feature Importance}
\label{sec:feat_imp}
The LightGBM framework \cite{ke2017lightgbm}, roughly speaking, builds a strong learner upon an ensemble of decision trees as weak learners. Decision trees assign at each learning step a dichotomous split on feature values based on the maximum obtainable information gain, so that it is a highly interpretable ML model. 

Looking at table \ref{table:feature_importance} we can see that features having the highest importance are common in almost all the stages, but each stage is characterized by a different ranking in the features' importance, confirming the differences in the various typology of classes and the necessity to use different models in each stage of the pipeline. Among the most frequent features we note the virtual size, that is known to be a highly discriminative feature in malware classification~\cite{sans2012}.

It is interesting to note that the most important features for the quarantine stage, that is, the last stage in which we try to recover some wrong classifications of the previous stages, belong to the major groups of the byte entropy and printable strings. Since it has been observed that entropy and the presence or not of readable strings are correlated to the presence of packed or encrypted code, this could be suggestive of the fact that the ``hardest" samples to correctly classify might be packed or encrypted, a known evasion techniques and a clear limitation to an approach based solely on static analysis. 

\begin{table}[t]
	\centering
	\begin{tabular}{ l | l }
		\hline
		Detection & Class (Type Threat)\\
		\hline
		Section:Entropy Hashed	&  Header:dll charateristics Hashed \\
		Data Directories:RESOURCE TABLE:size & Section:Entropy Hashed \\
		General Info:Vsize & Header:timestamp  \\
		Section:Vsize Hashed & General Info:Vsize \\
		Data Directories:RESOURCE TABLE:size & Data Directories:RESOURCE TABLE:size\\
		\hline
	\end{tabular}
	\label{tbl:top-5_type_dtc}
\end{table}
\begin{table}[h!]
	\centering
	\label{table_ASME_1}
	\begin{tabular}{ l | l }
		\hline
		Family	& Behavior \\
		\hline
		Header:dll charateristics Hashed& General Info:Vsize \\
		Data Directories:RESOURCE TABLE:size&Header:dll charateristics Hashed \\
		General Info:Vsize & Header:timestamp \\
		Section:Entropy Hashed & Data Directories:RESOURCE TABLE:virtual address \\
		Strings:printabledist & General Info: n° imports\\
		\hline
	\end{tabular}
	\label{tbl:top-5_fam_dtc}
\end{table}
\begin{table}[h!]
	\centering
	\label{table_ASME_2}
	\begin{tabular}{ l }
		\hline
		Quarantine \\
		\hline
		Byte Histogram \\
		Strings:printabledist \\
		Byte Entropy Histogram\\
		\hline
	\end{tabular}
	\caption{Top feature groups for the detection, threat-type, family, behaviour, and quarantine classification stages.}
	\label{table:feature_importance}
\end{table}

\subsection{Model Interpretability}
\label{sec:interpretability}

\begin{figure}[b!]
	\centering
	\begin{minipage}{0.45\textwidth}
		\centering
		\includegraphics[width=0.9\textwidth]{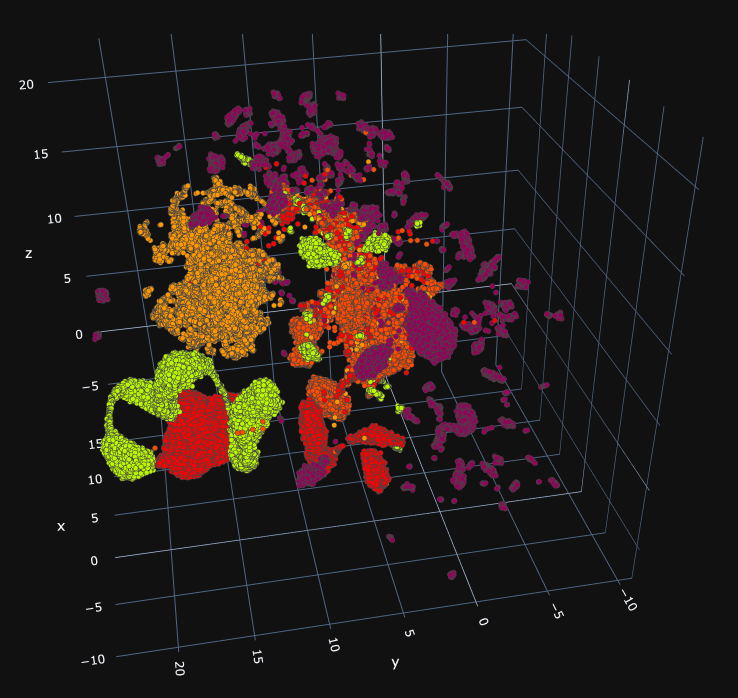}
	\end{minipage}\hfill
	\begin{minipage}{0.45\textwidth}
		\centering
		\includegraphics[width=0.9\textwidth]{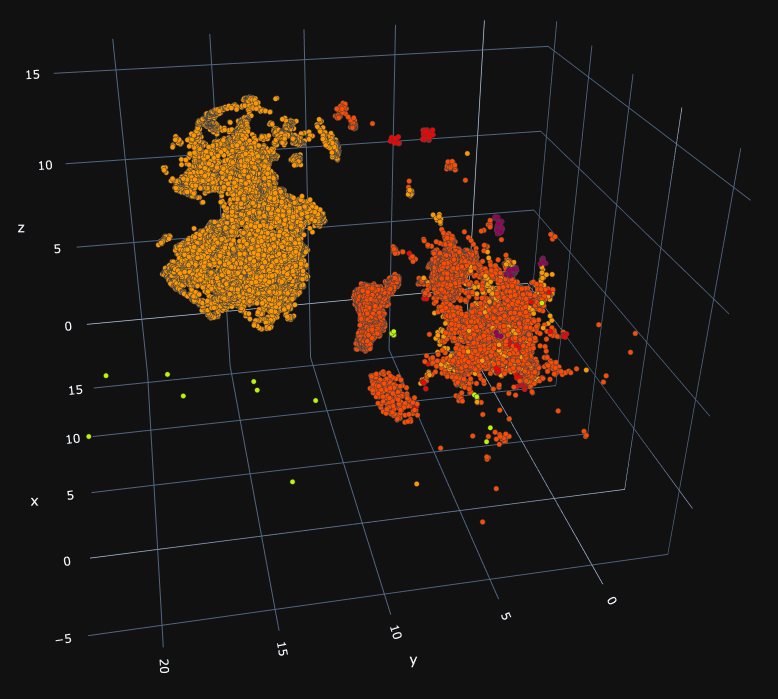}
	\end{minipage}\hfill
	\caption{3-d representation of 5 different malware families. On the left the train set (fitting phase), on the right the test set. The frequency of the classes in the train e test set are highly uneven.}
	\label{fig:3d_cluster_cm}
\end{figure}

In order to better understand the distribution of our data in the high-dimensional original feature space and, hence, be able to identify regions in which a certain classification stage might be less reliable, we apply UMAP~\cite{mcinnes2018-arXiv, mcinnes2018-software}, a non-linear and unsupervised dimensionality reduction technique, to the full feature vector of size $1252$ keeping the first $3$ components.
In turn, this allows also to suggest a potential data mislabelling that could need an analyst further inspection.

In figure \ref{fig:3d_cluster_cm}, we show the resulting embedding for the first five most frequent families.
We can see that all different families are generally well separated, often forming isolated sub-clusters that can be linked to specific types or variants of the main malware class. Other regions show instead a mixing of various classes. That is where the classification is more difficult and the model is more likely to fail. There are various reasons for the appearance of this regions: first of all, as explained in section \ref{sec:labelling}, the labels are not uniquely defined and thus it could simply be an effect of an imperfect labelling. Moreover, as pointed out in Section~\ref{sec:feat_imp}, many classification errors are likely to be linked to the presence of packing or encryption, that is a well known limitation to static analysis. Nevertheless, an analyst using our model could easily assess the confidence in the classification based on the position of the considered sample in this low-dimensional space, providing an useful tool for interpreting the results.

\section{Discussion}
\label{sec:discussion}

As discussed in~\ref{sec:pipe_acc}, some issues arise in the different stages of the pipeline due to: (\textit{i}) mislabeling of malware families reported in the EMBER dataset (refer to Section~\ref{sec:ember}), (\textit{ii}) overlaps among different threat types and behavior, and (\textit{iii}) few samples used to train classifiers in the last stages of the pipeline.

Regarding the second issue, we have observed that malware behaviors are not completely independent one from another: as an example, \textit{osmodify} is often confused with \textit{filemodify} and \textit{execdownload} behaviors, because in AVClass it is associated with rootkit malware that, in general, establishes communications with command-and-control servers to receive new commands and malicious payloads.
Finally, Table~\ref{tbl:pipe_acc} reports how the number of input samples, useful to train stage classifiers, is halved at each stage. As mentioned in Section~\ref{sec:pipe_acc}, few samples explain poorer performance in accuracy for family and behavior classifications.
\begin{figure}[tp]
	\centering
	\begin{minipage}{0.85\textwidth}
		\centering
		\includegraphics[width=0.9\textwidth]{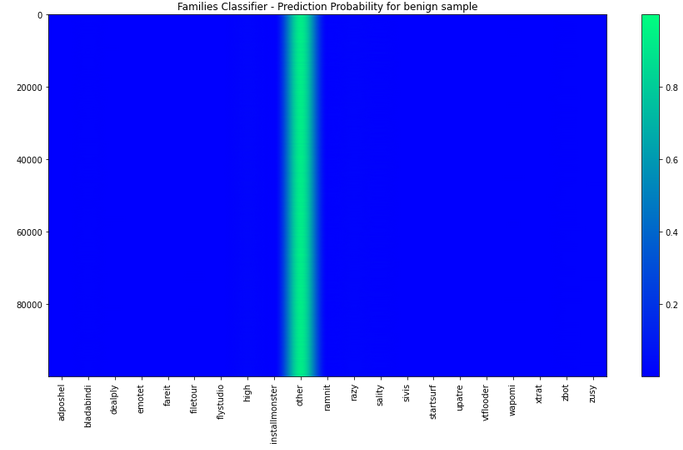}
	\end{minipage}\hfill
	\caption{Prediction probability obtained by providing in input previously unseen benign samples to a family classifier, trained only on malware samples.}
	\label{fig:benign_cm}
\end{figure}

A final consideration regards, instead, the choice of using classifiers able to discriminate between malware and benign samples to refine the overall classification process.
The idea was born out of trained classifiers' capacity of accurately distinguishing between malicious and benign samples.
As an example, Figure~\ref{fig:benign_cm} shows the confidence that a different family classifier, trained only on malware samples, has in assigning previously unseen benign samples to \textit{other} class. As already discussed in Section~\ref{sec:pipe_acc}, it includes more than $2,800$ malware families that have too few samples to build a classifier able to identify them, as already mentioned in Section~\ref{sec:clf_targ}.
It is worth noting that no benign samples has been classified as belonging to one of the most-frequent malware families of the EMBER dataset, listed on the \textit{x}-axis of Figure~\ref{fig:benign_cm}. The plot has been computed on $100,000$ benign samples, extracted from the test set used for the family classification stage.

This supports our hypothesis that the poor performance in correctly detecting benign samples in the classification stages of the pipeline is due to the very small number of benign samples during model training ($\approx1\%$ of all the benign samples).

\section{Conclusion and Future Work}
In this work we proposed a first implementation of a pipeline for a complete classification of Windows PE files using a machine learning approach on static features. The pipeline is able to separate benign and malicious samples, and for those samples classified as malicious it provides a exhaustive classification in terms of threat-type, malware family, and behaviour. Classification results, although suffering from known limitations such as the size of the training data, the imperfect labelling of the ground truth, and the semantic gap of models based on static features  only, are comparable to the current state of the art for similar works while providing much more detailed information on malware characteristics. Finally, the extracted feature vector characterising the raw PE and the specific ML model implemented provide an interpretable result, and the pipeline is scalable to much larger datasets. Therefore, we consider this work as a first step towards a useful tool that can help security analysts to manage novel threats, reducing time and costs of the analysis. For the near future, we plan to improve the described pipeline by: (\textit{i}) considering a larger dataset for training (\textit{ii}) fixing the ground truth and considering the possibility of a multi-label classification scheme, and (\textit{iii}) further explore the properties of the embedded features' space described briefly in section \ref{sec:interpretability}. It could be interesting to include also a detector for packed/encrypted samples in the early stages of the pipeline, and move to a hybrid approach, combining static and dynamic features for a better characterization of the sample files.

\label{sec:bib}
\bibliographystyle{unsrt}
\bibliography{references}
\end{document}